# Controlling the nonlinear relaxation of quantized propagating magnons in nanodevices

M. Mohseni,[1,*] Q. Wang,[2] B. Heinz,[1,3] M. Kewenig,[1] M. Schneider,[1] F. Kohl,[1] B. Lägel,[4] C. Dubs,[5] A. V. Chumak,[2] and P. Pirro[1]

[1] Fachbereich Physik and Landesforschungszentrum OPTIMAS, Technische Universität Kaiserslautern, 67663 Kaiserslautern, Germany

[2] Faculty of Physics, University of Vienna, Boltzmanngasse 5, A-1090 Vienna, Austria

[3] Graduate School Materials Science in Mainz, Staudingerweg 9, 55128 Mainz, Germany

[4] Nano Structuring Center, Technische Universität Kaiserslautern, 67663 Kaiserslautern, Germany

[5] INNOVENT e.V., Technologieentwicklung, Prüssingstraße 27B, 07745 Jena, Germany

Relaxation of linear magnetization dynamics is well described by the viscous Gilbert damping processes. However, for strong excitations, nonlinear damping processes such as the decay via magnon-magnon interactions emerge and trigger additional relaxation channels. Here, we use space- and time-resolved micro-focused Brillouin light scattering spectroscopy and micromagnetic simulations to investigate the nonlinear relaxation of strongly driven propagating spin-waves in yttrium iron garnet nanoconduits. We show that the nonlinear magnon relaxation in this highly quantized system possesses intermodal features, i.e. magnons scatter to higher-order quantized modes through a cascade of scattering events. We further show how to control such intermodal dissipation processes by quantization of the magnon band in single-mode devices, where this phenomenon approaches its fundamental limit. Our study extends the knowledge about nonlinear propagating spin-waves in nanostructures which is essential for the construction of advanced spin-wave elements as well as the realization of Bose-Einstein condensates in scaled systems.

Relaxation of magnons, the quanta of spin waves (SWs), due to magnetic damping is a complicated process and involves different (non)linear contributions. Relaxation mechanisms which can be described by the phenomenological Gilbert damping drive the magnetization towards its equilibrium state by e.g. dissipating the energy to the lattice. It is one of the key elements of performance in many practical devices and fundamental phenomena [1–10].

Dissipation of the energy can be more intricate for strongly driven excitations, where nonlinear relaxation mechanisms via magnon-magnon interactions open up additional dissipation channels [11–17]. Unlike the Gilbert damping, these types of intrinsic dissipation processes can redistribute the magnon energy within the magnon spectrum [18–27].

The classical works of Suhl predicted that large amplitude uniform magnetization oscillations lead to the onset of instability processes, allowing the nonlinear relaxation of strongly driven magnons by a decay into secondary magnon modes [25]. In particular, the common second-order Suhl instability process can be: (*i*) a disadvantage since it comes along with detrimental influence on the magnon transport and decay characteristics, potentially dominating the competing linear damping [17,22,28], or, (*ii*) an advantage by providing additional degrees of freedom of magnon transport for device architectures and quantum computing concepts [23,29,30]. So far, most of such investigations in scaled systems, which are of large interest for applications, have been carried out for standing SW modes with vanishing momentum ($k = 0$), e.g. the Ferromagnetic resonance (FMR) mode. However, SWs carrying a momentum are not only essential for applications, but they possess an enriched physics behind their nonlinear instabilities due to the increased amount of potential scattering channels. Nevertheless, little investigations have been carried out in this direction yet.

Recent development of ultra-low damping nanoscale systems based on YIG, the most promising hosts for SWs, provides access to quasi-1D systems with highly quantized magnon spectra [31,32]. By imposing limitations on the available relaxation channels due to the strong quantization of the magnon band, and a drastically modified SW characteristics including the SW dispersion relation, mode profile and their ellipticity, nonlinear SW dynamics in such devices can be different compared to continuous films and quasi-2D systems [33-34]. Furthermore, recent experimental and theoretical studies of SW dynamics and magnon condensates in nanoscopic systems [32, 35,36] enforce us to better understand nonlinear SW dynamics and magnon thermalization processes in nano-scaled 1D systems.

Here, we use space- and time-resolved micro-focused Brillouin light scattering (µBLS) to uncover the mechanism of nonlinear relaxation of strongly driven propagating magnons via the second-order Suhl instability in YIG nanoconduits. We demonstrate how magnons nonlinearly relax to other quantized modes via four-magnon scattering processes, and such nonlinear processes can be controlled using quantization of the magnon band.

To demonstrate the effect of quantization on the nonlinear dynamics, we use two exemplary magnonic nanoconduits structured from a Liquid Phase Epitaxial (LPE) YIG film grown on top of a Gadolinium Gallium Garnet (GGG) substrate [37]. The multi-mode nanoconduit with a lateral width of $w = $ 400 nm (Fig. 1a) and a thickness of $d = $ 85 nm was fabricated using a hard mask and ion beam milling process [31]. A comparative single-mode conduit with a smaller width of $w = $ 100 nm and $d = $ 44 nm was fabricated using a similar method (Fig. 1b). SWs in both devices are excited by a microwave antenna which is placed on top of the nanoconduits by electron beam lithography and a lift-off process [31]. Applying a microwave $rf$ current to the antenna generates a dynamic Oersted field which in return excites SWs resonantly, see supplemental materials SM [40]. The detection of the generated SWs has been carried out using space- and time-resolved µBLS [38]. An incident laser light with an effective spot size of 300 nm (focused by a ×100 microscope objective with a numerical aperture NA=0.85) is used to probe the SWs through the GGG substrate under the antenna. The inelastically scattered light was analyzed using a tandem Fabry-Perot interferometer to obtain the frequency and intensity of the magnons.

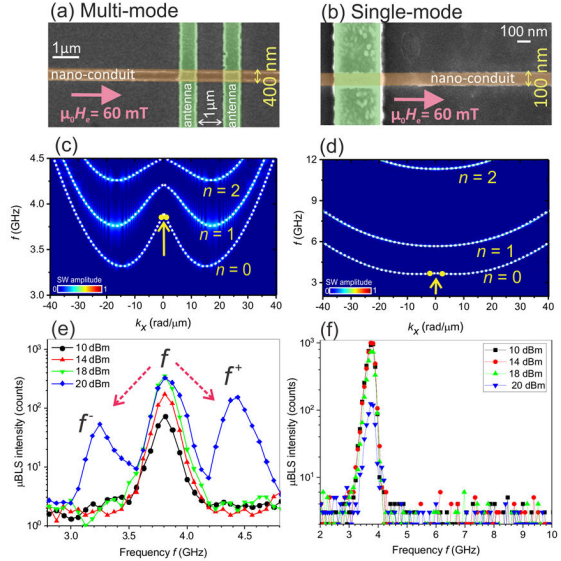

FIG. 1. (a)-(b) SEM images of the $w = $ 400 nm (multi-mode) and $w = $100 nm (single-mode) wide conduits (shaded in orange), respectively. (c)-(d) Magnon band structures of the multi-mode and single-mode conduits, respectively. Color plots are obtained by micromagnetic simulations and dashed lines from analytical calculations. Note the different scales of the frequencies. (e)-(f) Measured spin-wave spectra of the multi-mode and single-mode conduits in the presence of different powers, respectively. The excited modes are represented by the yellow dots in (c) and (d).

A static external field ($\mu_0 H_e$ = 60 mT) saturates the nanoconduits along their length. Thus, the wave vector of the propagating SWs is parallel to the magnetization vector, $k \parallel M$, and waveguide (WG) modes appears [32]. The width of the multi-mode waveguide is large enough to ensure dipolar pinning of the spins at the edges, while spins at the edges of the single-mode conduit are fully unpinned [32]. Moreover, due to the interplay between the contributions of the dipolar and exchange energy to the SW dispersion, the different WG modes are well quantized on the frequency axis. The dispersion relation of the fundamental mode and the first two WG modes are shown in Fig 1c-d, in which the dashed lines are analytical results based on method discusses in Ref [32], and the color plot is obtained by micromagnetic simulations using the MuMax 3.0 package [39, 40]. The fundamental mode and higher order WG modes are labeled as $n = 0$ and $n = 1, 2$ respectively. Please note that the spectrum is much more dilute in the 100 nm wide conduit due to the higher contribution of the exchange energy to the magnon band structure, which leads to a strong quantization and the absence of degenerate states among modes (single-mode system for wave vectors below approx. 40 rad/μm).

We first set the *rf* frequency to $f = 3.85$ GHz where dipolar SWs having a wave vector of $k_x = 1.5$ rad/μm are excited in the multi-mode device [40]. To characterize the linear SW dynamics, we set the *rf* power to $P = 10$ dBm and measure the intensity of the generated magnons as displayed in Fig 1e (black circles). Up to $P = 18$ dBm, only the frequency of the resonantly driven SW mode is observed (red and green triangles). A further increase in the *rf* power up to $P = 20$ dBm (blue curve) leads to the appearance of two additional peaks in the SW frequency spectrum labeled as $f^-$ and $f^+$ in Fig. 1e. We refer to these magnons as *secondary magnons* which are modes populated by nonlinear scattering processes. They have the lowest threshold for the observed instability process and can fulfill the fundamental conservation laws to permit the scattering process [22]. The energy and momentum conservation laws of these processes generally read [18,20,22,28],

$$f^1 + f^2 = f^3 + f^4, \quad \boldsymbol{k}^1 + \boldsymbol{k}^2 = \boldsymbol{k}^3 + \boldsymbol{k}^4 \quad (1)$$

where two magnons with the frequencies $f^1$ & $f^2$ and momenta $\boldsymbol{k}^1$ & $\boldsymbol{k}^2$ scatter to two magnons with the frequencies $f^3$ & $f^4$ and momenta $\boldsymbol{k}^3$ & $\boldsymbol{k}^4$. Note that the lateral component of the $\boldsymbol{k}$ vector is symmetric, and the out of plane component is zero in this frequency range due to the small thickness. In our experiments, two magnons with a frequency of $f = 3.85$ GHz scatter finally to two magnons with the frequencies of $f^3 = f^- = 3.25$ GHz and $f^4 = f^+ = 4.45$ GHz. We note that this process is not a special peculiarity of the chosen spectral position, see SM [40].

For comparison, we now investigate the same nonlinear process in the comparative single-mode waveguide. We set the $f = 3.71$ GHz and measure the intensity of the driven mode as shown in Fig. 1f. Clearly, even in the presence of high powers like $P = 20$ dBm, side peaks cannot be observed, evidencing the absence of a similar nonlinear dissipation processes. Here, only the μBLS intensity drops at high powers which is caused by the nonlinear frequency shift of the dispersion relation and possible impacts of the higher temperature [31, 41]. In principle, the absence of side peaks demonstrates that such scattering processes can be efficiently suppressed in narrower conduits where the magnon band structure is highly quantized and therefore, the fundamental conservation laws required for the scattering processes cannot be fulfilled.

To understand the fundamental differences between the two waveguide types, let us investigate the observed nonlinear dynamics in the multi-mode conduit in more detail. A nonlinear scattering instability is characterized by a clear threshold of the initial magnon intensity which is required for its onset [18,22,24,42]. Neglecting SW radiation losses, the threshold magnon amplitude is defined by the effective relaxation frequency of the secondary magnons divided by the four-magnon coupling strength [16,22]. To investigate the threshold behavior in the multi-mode conduit in which the scattering is observed, we sweep the *rf* power for a fixed frequency $f = 3.85$ GHz as shown in Fig. 2. Once the instability threshold is reached at $P = 18$ dBm (indicated by the black arrow), the growth rate of the directly excited magnon intensity as a function of microwave power drops. Increasing the power to $P = 19$ dBm leads to an abrupt increase of the intensity of the secondary magnons labeled as $f^+$ and $f^-$ (indicated by the gray arrow). From this power ($P = 19$ dBm) on, the intensity growth rate of the directly excited mode with respect to the power is decreased, evidencing that the energy transfers to the secondary magnon modes.

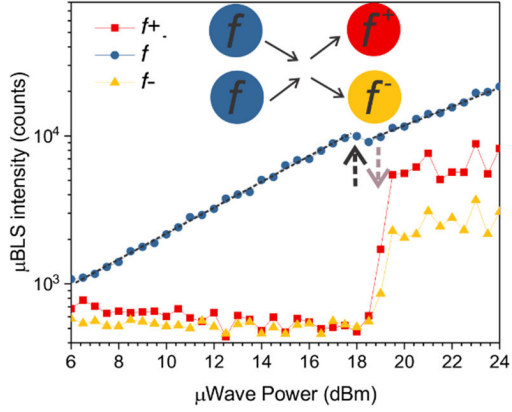

FIG. 2. Spin-wave amplitude in the multi-mode conduit as a function of microwave excitation power when $f$ = 3.85 GHz. The secondary magnons created by the second order Suhl instability are denoted as $f^+$ and $f^-$. The back and gray arrows indicate the onset of instability and the rise of the secondary magnons, respectively.

A closer look on Fig. 2 near the instability threshold opens the question what happens when the instability threshold is approached at $P$ = 18 dBm and the μBLS intensity of the initially excited mode drops, while the amplitudes of the secondary magnons at $f^+$ and $f^-$ are still at the thermal level, implying the absence of magnon scattering to these modes.

We perform micromagnetic simulations to uncover the wave vector of the scattered magnons and address the discussed question. Figure 3a shows the frequency spectrum of the simulated multi-mode conduit ($f$ = 3.85 GHz) in which different amplitude of the $rf$ currents are used to drive the system. For a small $rf$ current equal to $i_{rf}$ = 4 mA, only the resonantly excited SWs can be observed in the frequency spectrum (black curve). The corresponding population of the magnon band is depicted in Fig. 3b showing the wave vector of $k_x$ = 1.5 rad/μm of the directly excited mode. Increasing the $rf$ current to a higher value of $i_{rf}$ = 8 mA increases the amplitude and the linewidth of the resonant SWs (red curve in Fig 3.a) [28]. As shown in Fig. 3c, this is related to the onset of a first-level four magnon scattering process in which the frequency of the magnons is conserved. Such a process cannot be observed in the measured frequency spectrum of the conduit, but it can manifest itself in the observed drop of the directly excited mode intensity with power. As evidenced by the simulations, two incoming magnons from the resonantly driven mode with opposite momenta scatter to two outgoing magnons at the same frequency, but with different momenta. The scattered magnons populate the fundamental mode ($n$ = 0) at a higher wave number of $k_x$ = 30 $rad$/μm, and two spectral position at the first WG mode ($n$ = 1). These *frequency-conserving* scattering processes which are similar to plane films [25] are indicated by the pink arrows in Fig. 3c, and can also be observed in the single-mode conduit, see SM [40].

A further increase of the $rf$ current to $i_{rf}$ = 13 mA leads to the onset of the sideband peaks in the frequency spectrum (blue curve in Fig 3.a), similar to the experiments. As evidenced from the simulated band structure (Fig. 3d), this is due to the second level of the magnon scattering cascade. Once the magnons scattered by the first level process to the $n$=1 WG mode reach a critical amplitude, they undergo themselves another second order instability. In this process, two magnons with the frequency of $f$ = 3.85 GHz and identical momentum of $k_x$ = 10.7 $rad$/μm at the first WG mode ($n$ = 1), scatter to two outgoing magnons with the frequencies of $f^-$ = 3.46 GHz and $f^+$ = 4.24 GHz at the fundamental mode ($n$ = 0) and the second WG mode ($n$ = 2), respectively. The simulated values are in very good agreement with the experimentally obtained frequencies.

In Fig. 3d, this type of *frequency-nonconserving* scattering is represented by the red arrows. The scattered magnons feature $k_x^+ = 14.3\ rad/μm$ and $k_x^- = 7.1\ rad/μm$, assuring momentum conservation laws given by $2k_x = k_x^+ + k_x^-$. We note that the second scattering step clearly shows that the finite momentum of the ingoing magnon opens the opportunity to scatter to two new, different frequencies and thus, to redistribute the magnon energy towards the bottom of the spectrum and to higher frequencies (modes). Unlike the first level process, it involves only magnons of a single propagation direction (+$k$ or –$k$) and can only occur for propagating waves.

This is evidenced by the momentum and energy conservation laws which require a finite sum of the momenta of the two incoming magnons to allow for a frequency non-degenerated splitting. This is a significant difference to the nonlinear instabilities of the FMR mode without momentum ($k_x = 0$) in which magnon instabilities are always degenerated [43-44]. Thus, if the FMR undergoes a second-order instability, this process never leads to a redistribution of the magnon energy across the spectrum.

(Fig. 3c) cannot be detected experimentally due to the maximum detectable momentum using μBLS spectroscopy, which is approximately $k_x \sim 21$ rad/μm in our experiments [38]. This explains at least partially the decrease of the measured magnon intensity at the driving frequency. Since the different levels of the cascade process have different threshold powers, the nonlinear scattering to the secondary magnon modes at different frequencies is observed at a slightly higher power than the start of the drop of the intensity at the directly excited frequency. In addition, the limited wave vector sensitivity of the BLS can pose inconsistency for the SW amplitude observed in the simulations compared to the experiments.

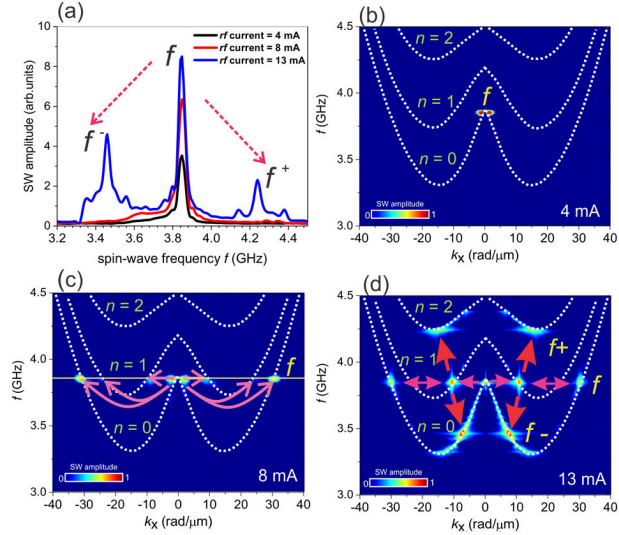

FIG. 3. Results of the micromagnetic simulations in the multi-mode conduit. (a) spin-wave frequency spectra when the microwave current varies. (b-d) Magnon band structures (linear scale) of the driven system correspond to the black, red and blue curves in (a), respectively. The scaling of b-d is independent from each other.

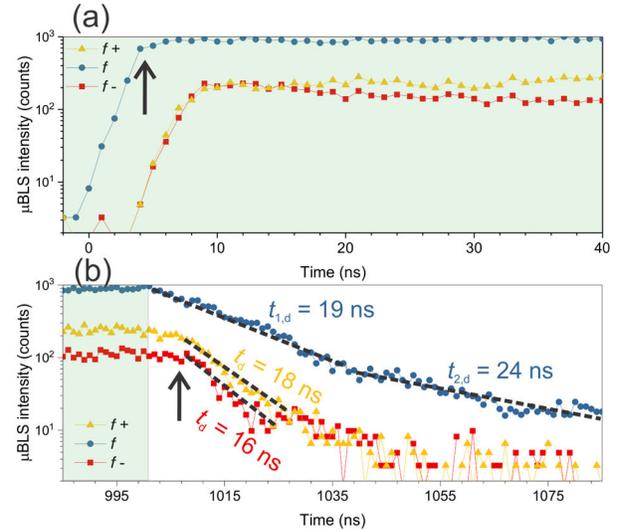

FIG. 4. Time-resolved spin-wave amplitude measured by μBLS spectroscopy. (a) Beginning of the pulse. (b) End of the pulse. Black arrows indicate the onset and the decay of the instability, respectively. Note that the decay rates correspond to the intensity of the magnons.

The properties of the cascade-like magnon scattering events coupling different waveguide modes in the multi-mode waveguide and the absence of this effect in the single mode waveguide also implies that thermalization of magnons is significantly changed in systems with strongly diluted spectra compared to earlier investigations in systems which quasi-continuous spectra.

The simulations also explain the observed peculiarity in the threshold curve of the experiments as were discussed in the context of Fig. 2. Indeed, the magnons scattered to higher wave numbers via the first level frequency-conserving scattering process

To further characterize the impact of the nonlinear relaxation on the total relaxation of the system [16], we perform time-resolved μBLS measurements in the multi-mode conduit. The measured intensity of the driven and secondary magnons at the beginning and the end of a 1μs long microwave *rf* pulse ($f$ = 3.85 GHz and $P$ = 24 dBm) at the measurement position are shown in Fig. 4a-b. Figure 4a illustrates that the resonantly driven SW mode (blue curve) undergoes the second-level four-magnon scattering after $t \sim 4$ ns, evidenced by the rise of the secondary magnons (yellow and red curves). This is indicated by the black arrow in Fig. 4a. Note that the growth rate of the driven mode drops immediately when the rise of the secondary magnons sets in, evidencing the conservation of the energy in the nonlinear redistribution process.

The decay of the magnons at the end of pulse is presented in Fig. 4b. In particular, the decay of the secondary magnons begins once the intensity of the driven SW mode is decayed enough after $t \sim 4$ ns (indicated by the black arrow). More interestingly, the decay of the magnons at the resonantly driven frequency to the thermal level includes two steps manifesting the high nonlinearity of the dynamics. First, it decays with an exponential decay time of $t_{1,d} = 19$ ns, which is accompanied by the decay of the secondary magnons at $f+$ and $f-$. Afterwards, it decays with a longer exponential decay time of $t_{2,d} = 24$ ns suggesting a transition from a nonlinear relaxation to a linear relaxation with a lower decay rate. In other words, the first decay includes an energy flow to the secondary magnons which acts as an additional dissipation channel for the driven magnons. After the secondary magnons decayed to the thermal level, this additional dissipation channel is switched off, which leads to a slower decay time of the driven SWs.

In summary, we explored the nonlinear relaxation of strongly driven propagating spin waves in nanodevices. The finite momentum of the magnons investigated in our study provides an additional playground for the nonlinear magnon instability processes. Furthermore, it was shown that such intermodal dissipation process is strongly suppressed in systems with a strongly quantized magnon band (single-mode systems), suggesting the fundamental limitation of this process in nanodevices. This can open a new avenue for coherent nonlinear nano-magnonics. The nonlinear dynamics studied in this letter are general and thus, can be applied to devices based on other deposition techniques as well. Our study can be used for several device architectures, namely, frequency mixers [45], squeezed states [46], signal and data processing units [29, 47-50], and quantum computing concepts [23], and further open doors to engineered dissipation of magnons in nanodevices.

Acknowledgments

The authors thank Burkard Hillebrands for support and valuable discussions. This project is funded by the Deutsche Forschungsgemeinschaft (DFG, German Research Foundation) - TRR 173 - 268565370 ("Spin+X", Project B01) and by the project - 271741898, the European Research Council within the Starting Grant No. 678309 "MagnonCircuits" and by the Austrian Science Fund (FWF) through the project I 4696-N. We appreciate our colleagues from the Nano Structuring Center of the TU Kaiserslautern for their assistance in sample preparation. B.H. acknowledges support by the Graduate School Material Science in Mainz (MAINZ).

Correspondence to: *mohseni@rhrk.uni-kl.de